\documentclass[pra,twocolumn,showpacs,amsmath,amssymb,superscriptaddress]{revtex4}

\usepackage{graphicx}
\usepackage{dcolumn}
\usepackage{bm}

\begin{document}

\title{
Controlled pairing symmetry of the superfluid state in systems of three-component repulsive fermionic atoms in optical lattices
}

\author{Sei-ichiro Suga}%
\affiliation{Graduate School of Engineering, University of Hyogo, Himeji 671-2280, Japan}


\date{\today}

\begin{abstract}
We investigate the pairing symmetry of the superfluid state in repulsively interacting three-component (color) fermionic atoms in optical lattices.
When two of the three color-dependent repulsions are much stronger than the other, pairing symmetry is an extended $s$ wave, although the superfluid state appears adjacent to the paired Mott insulator in the phase diagram. 
On the other hand, when two of the three color-dependent repulsions are weaker than the other, pairing symmetry is a $d_{x^2-y^2}$-wave.
This change in pairing symmetry is attributed to the change in the dominant quantum fluctuations from the density fluctuations of unpaired atoms and the color-density wave fluctuations to the color-selective antiferromagnet fluctuations. This phenomenon can be studied using existing experimental techniques.
\end{abstract}


\preprint{APS/123-QED}

\maketitle

\section{\label{sec:level1}Introduction}

The cold atoms in optical lattices are well known for their high controllability.
Recently, a three-component (color degrees of freedom) fermionic gas has been realized using $^6{\rm Li}$ atoms with three hyperfine states \cite{Ottenstein2008,Huckans2009}. In this system a balanced population of each color atoms has been achieved, while color-dependent three interactions are different, preventing the emergence of SU(3) symmetry. Highly symmetric systems with SU(6) symmetry and SU(2)$\times$SU(6) symmetry have been realized in $^{173}$Yb atoms \cite{Fukuhara2007,Taie2012} and in a mixture of $^{171}$Yb atoms and $^{173}$Yb atoms \cite{Taie2010}, respectively.  
A SU(10) symmetric system has been realized in $^{87}$Sr atoms \cite{DeSalvo2010}. 
These results mean that the cold fermionic atoms in optical lattices constitute a quantum simulator for studying the correlation effects of multicomponent lattice fermions.

Currently, one of the most attractive research fields in condensed-matter physics is superconductivity in strongly correlated multiorbital systems. For instance, superconductivity mediated by orbital fluctuations has been discussed as a possible mechanism of unconventional characters of the heavy-fermion superconductivity in $\rm PrOs_{4}Sb_{12}$ \cite{Bauer2002,Goremychki2004,Kuwahara2005} and $\rm PrTr_2Al_{20}  (Tr=Ti, V)$ \cite{Sakai2012,Matsubayashi2012,Tsujimoto2014}.  
For iron-based superconductors, it has been pointed out that the electron correlation effects in the multiorbital system play a key role in the appearance of superconductivity. As regards Cooper-pairing symmetry, both the spin fluctuation-mediated $s_{\pm}$ wave \cite{Kuroki2008,Mazin2008,Wang2009,Hirschfeld2011,Chubukov2012} and the orbital fluctuation-mediated $s_{++}$ wave \cite{Kontani2010,Yanagi2010,Onari2012} have been proposed theoretically. In the former the gap function changes the sign between different Fermi surfaces, while in the latter the sign of the gap function is unchanged. 
 Pairing symmetry and its origin have been actively debated.
As discussed later, in three-component repulsive fermionic atoms in optical lattices  
three kinds of quantum fluctuations can be controlled, leading to a change in the superfluid pairing symmetry between an $s$ wave and a $d$ wave.   
Thus, three-component repulsive fermionic atoms in optical lattices act as a minimal model for discussing the  Cooper-pairing mechanism and symmetry of multicomponent repulsive lattice fermions.  
Our findings provide an insight into the Cooper-pairing mechanism of, for instance, multiorbital correlated electron systems.

We have already investigated repulsively interacting three-component fermionic atoms in optical lattices with a balanced population \cite{Miyatake2010,Inaba2010b,Miyatake2010b,Inaba2012,Inaba2013}.
We have shown that characteristic Mott states can appear in spite of half filling, where the average atom number per site is the noninteger 3/2. When two of the three repulsions are stronger than the other, pairs of weakly repulsing atoms are formed to avoid the two stronger repulsions. This Mott state is called the paired Mott insulator (PMI) \cite{Inaba2010b}. When two of the three repulsions are weaker than the other, atoms with the strongest repulsion undergo a Mott transition while atoms with the remaining color stay a Fermi liquid. This state is called the color-selective Mott state (CSM) \cite{Miyatake2010,Inaba2010b}.  
In the ground state at half filling, two kinds of the staggered ordered states appear in the respective parameter regions: a color-density wave (CDW) and a color-selective antiferromagnet (CSAF) \cite{Miyatake2010}. In the CDW paired atoms with two different colors and atoms with the remaining color alternately occupy different sites, while in the CSAF localized atoms with two different colors alternately occupy different sites. 
These results suggest a possibility of emergence of a superfluid state. 
Using a dynamical mean-field theory (DMFT) we have shown that a superfluid state appears at close to half filling in the parameters for the CDW \cite{Inaba2012,Inaba2013}. It has been also shown that a superfluid appears at close to the PMI transition point in the `color paramagnetic' sector at half filling \cite{Inaba2012,Inaba2013,Koga2014}. 
In these superfluid states, atoms with the weakest repulsion form Cooper pairs while atoms with the remaining color stay a Fermi liquid. 
These calculations are based on a DMFT, thus making it difficult to discuss Cooper-pairing symmetry.

In this paper, we investigate superfluid pairing symmetry in repulsively interacting three-component fermionic atoms in square optical lattices by using an \'{E}liashberg equation.
We show that when two of the three repulsions are sufficiently strong, pairing symmetry is an extended $s$ wave, although the superfluid state appears adjacent to the PMI in the phase diagram. 
When two of the three repulsions are weaker than the other, a superfluid state with a $d_{x^2-y^2}$-wave pairing appears. 
From our calculations of the effective interaction, we show that this change in pairing symmetry can be attributed to the change in the dominant quantum fluctuations among the local density fluctuations of unpaired atoms, the nonlocal CDW fluctuations, and the  nonlocal CSAF fluctuations.

\section{\label{sec:level2}Model and Method}
The low-energy properties of the ultracold atoms in optical lattices are well described by the following Hubbard-type Hamiltonian \cite{Jaksch,Lewenstein}:
\begin{eqnarray}
\hat{\cal H}=&-&t \sum_{\langle i,j \rangle}\sum_{\alpha=1}^{3}
       \hat{a}^\dag_{i\alpha} \hat{a}_{j\alpha}
  - \sum_{i}\sum_{\alpha=1}^{3} \mu_\alpha \hat{n}_{i \alpha} \nonumber \\
  &+& \frac{1}{2}\sum_{i}\sum_{\alpha\not=\beta}
       U_{\alpha\beta} \hat{n}_{i \alpha} \hat{n}_{i \beta},
\label{model}
\end{eqnarray}
where $t$ is the nearest-neighbor hopping integral, and $\hat{a}^\dag_{i\alpha} (\hat{a}_{i\alpha})$ and $\hat{n}_{i\alpha}$ are creation (annihilation) and number operators of a fermion with color $\alpha(=1,2,3)$ at the $i$th site. The subscript $\langle i,j \rangle$ is the summation over the nearest-neighbor sites.
Filling $N$ is given by $N=\sum_\alpha n_\alpha$, where $n_\alpha\equiv \langle \hat{n}_{i \alpha} \rangle$ is the average number of color-$\alpha$ atoms at a site.
We focus on the situation with a balanced population. Thus we set the chemical potential $\mu_{\alpha}$ so as to satisfy $n_1=n_2=n_3\equiv n$, which yields $N=3/2$ at half filling. 
The on-site repulsive interactions between color-$\alpha$ and $\beta$ atoms are denoted as $U_{\alpha\beta}$.
We consider that Cooper pairs are formed between color-1 and 2 atoms. For this purpose, we set $U_{12}\equiv U$ and $U_{23}=U_{31}\equiv U'$ for convenience, which yield $\mu_1=\mu_2\equiv\mu$ and $\mu_3\equiv\mu'$.

We discuss pairing symmetry by solving the following \'{E}liashberg equation for the superfluid order parameter $\Delta(\bm{k})$ within a weak-coupling theory:
\begin{eqnarray}
\lambda \Delta(\bm{k})=-\frac{1}{M}\sum_{\bm{k}'}\tilde{U}(\bm{k}-\bm{k}') \frac{\tanh(\beta \epsilon_{\bm{k}'}/2)}{2\epsilon_{\bm{k}'}}\Delta(\bm{k}'),
\label{eliash}
\end{eqnarray}
where $M$ is the number of $\bm{k}$-point meshes, $\beta=1/(k_{\rm B}T)$, and $\epsilon_{\bm{k}}$ is the bare energy dispersion of the color-1 and 2 atoms measured from the chemical potential.
The eigenvalue $\lambda$ is a measure of the dominant pairing symmetry and it reaches unity at the transition point.
The effective pairing interaction between color-1 and 2 atoms is calculated by collecting random-phase-approximation-type bubble diagrams and ladder-type diagrams, and is given as
\begin{eqnarray}
\tilde{U}(\bm{q})=U+\frac{3}{2}U^2 \chi_{\rm s}(\bm{q})-\frac{1}{2}U^2 \chi_{\rm c}(\bm{q}),
\label{effU}
\end{eqnarray}
where
\begin{eqnarray}
\chi_{\rm s}(\bm{q})=\frac{\chi^{}_{1}(\bm{q})}{1-U\chi^{}_{1}(\bm{q})},
\label{chis}
\end{eqnarray}
\begin{eqnarray}
\chi_{\rm c}(\bm{q})=
\frac{\chi^{}_{1}(\bm{q})+2(\frac{U'}{U})^2\chi^{}_{3}(\bm{q})-2U'\frac{U'}{U}\chi^{}_{1}(\bm{q})\chi^{}_{3}(\bm{q})}
{1+U\chi^{}_{1}(\bm{q})-2U'^2\chi^{}_{1}(\bm{q})\chi^{}_{3}(\bm{q})}.
\label{chic}
\end{eqnarray}
Here, $\chi^{}_{1}(\bm{q})=\chi^{}_{2}(\bm{q})=\frac{1}{M} \sum_{\bm{k}} \frac{f(\epsilon_{\bm{k+q}})-f(\epsilon_{\bm{k}})}{-\epsilon_{\bm{k+q}}+\epsilon_{\bm{k}}}$ is the bare susceptibility of color-1 and 2 atoms, $\chi^{}_{3}(\bm{q})=\frac{1}{M} \sum_{\bm{k}} \frac{f(\epsilon'_{\bm{k+q}})-f(\epsilon'_{\bm{k}})}{-\epsilon'_{\bm{k+q}}+\epsilon'_{\bm{k}}}$ is the bare susceptibility of color-3 atoms with $\epsilon'_{\bm{k}}$ being the bare energy dispersion of color-3 atoms measured from the chemical potential, and $f(x)$ is the Fermi distribution function.

According to  Eq. (\ref{chic}), $\chi_{\rm c}(\bm{q})$ has the possibility of showing a divergent peak because of the third term in the denominator. This result is characteristic of the three-component repulsive fermionic atoms in optical lattices. 
Actually, for $U'=0$ the third term in the denominator disappears and $\chi_c(\bm{q})$ is reduced to the charge susceptibility of the conventional two-component Hubbard model, where $\chi_{\rm c}(\bm{q})$ never diverges. 
On the other hand, $\chi_s(\bm{q})$ in Eq. (\ref{chis}) is the same form as the spin susceptibility of the conventional two-component Hubbard model.

We consider the system in square optical lattices. Thus, $\epsilon_{\bm{k}}=-2t\left[\cos(k_x)+\cos(k_y)\right]-\mu$ and $\epsilon'_{\bm{k}}=-2t\left[\cos(k_x)+\cos(k_y)\right]-\mu'$.
We employ $M=512\times512$ $\bm{k}$-point meshes in the numerical calculations.
We set $t$ in units of energy.

\section{\label{sec:level3}Results}
\subsection{Two stronger repulsions than the other}

\begin{figure}
\begin{center}
\includegraphics[width=\linewidth]{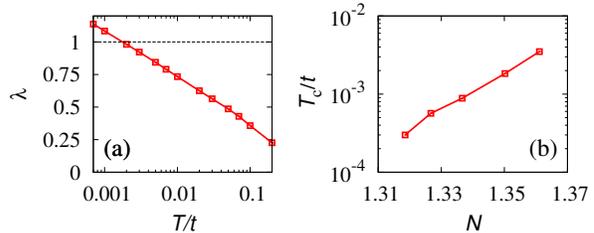}
\end{center}
\caption{(Color online) (a) Eigenvalue $\lambda$ as a function of temperature $T$ for $U/U'=0.01$, $U'/t=1.2$, and $N \sim1.35$.   (b) Transition temperature $T_{\rm c}$ as a function of filling $N$ for $U/U'=0.01$ and $U'/t=1.2$. 
}
\label{f1}
\end{figure}
\begin{figure}
\begin{center}
\includegraphics[width=\linewidth]{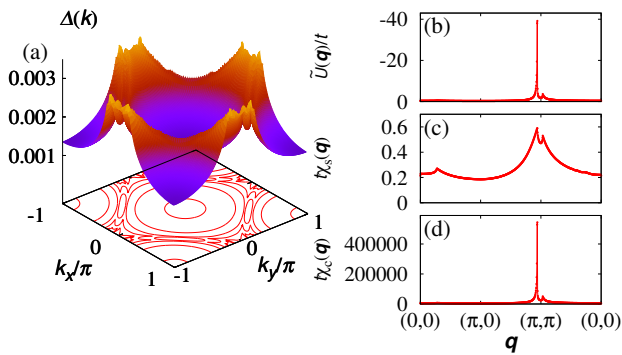}
\end{center}
\caption{(Color online) (a) Superfluid order parameter $\Delta(\bm{k})$, (b) effective pairing interaction $\tilde{U}(\bm{q})$, and susceptibilities (c) $\chi_s(\bm{q})$ and (d) $\chi_c(\bm{q})$ for $U/U'=0.01$, $U'/t=1.2$, $N \sim1.35$, and $T/t=0.003$,  which is slightly higher than $T_{\rm c}/t \sim 0.0018$. Momenta satisfy $\bm{q}=\bm{k}-\bm{k'}$. 
}
\label{f2}
\end{figure}
\begin{figure*}[htb]
\begin{center}
\includegraphics[width=130mm]{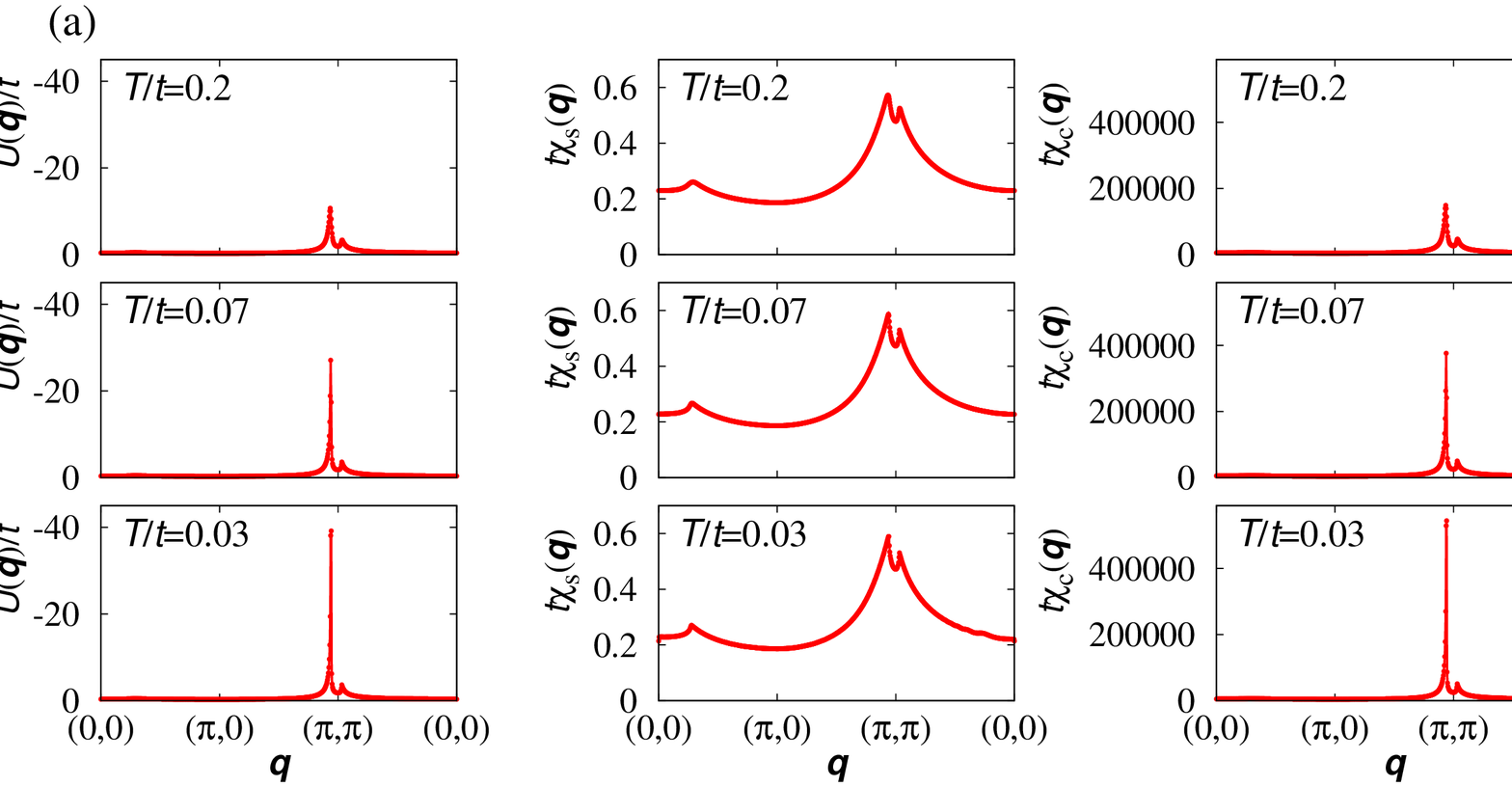}
\includegraphics[width=130mm]{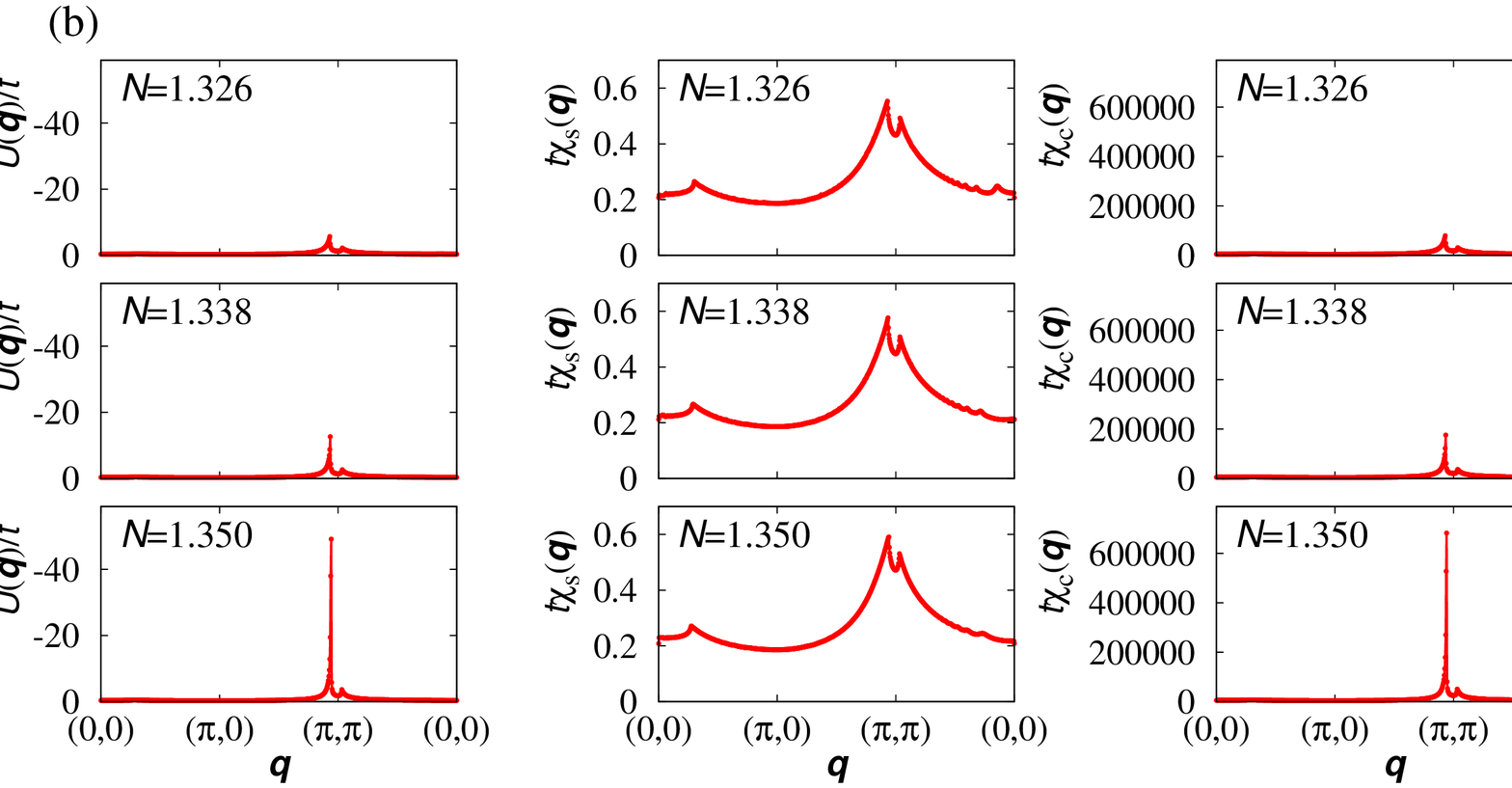}
\end{center}
\caption{(Color online) (a) Temperature dependencies of effective interaction $\tilde{U}(\bm{q})$, susceptibilities $\chi_s(\bm{q})$ and  $\chi_c(\bm{q})$ for $U/U'=0.01$, $U'/t=1.2$, and $N \sim1.35$. 
(b) Filling dependencies of effective interaction $\tilde{U}(\bm{q})$, susceptibilities $\chi_s(\bm{q})$ and  $\chi_c(\bm{q})$ for $U/U'=0.01$, $U'/t=1.2$, and $T\sim 1.2T_{\rm c}$. 
}
\label{f3}
\end{figure*}
In Fig. \ref{f1}(a), we show the temperature dependence of the eigenvalue $\lambda$ for $U'/t=1.2$, $U/U'=0.01$, and $N \sim1.35$. The parameters $U/U'=0.01$ and $N \sim1.35$ correspond to the condition for the appearance of the superfluid state in Ref. \cite{Inaba2012}. 
We find that $\lambda$ increases with decreasing temperature, resulting in the transition temperature $T_{\rm c}/t \sim 0.0018$.  
Figure \ref{f2}(a) shows the order parameter $\Delta(\bm{k})$ for $U'/t=1.2$, $U/U'=0.01$, and $N \sim1.35$ at $T/t=0.003$. We find that $\Delta(\bm{k})$ is fully gapped (nodeless) and has large amplitudes at $\bm{k} \sim (\pm\pi,0)$ and $(0,\pm\pi)$. The results indicate that pairing symmetry is an extended $s$ wave.
We discuss the origin of the large amplitude of $\Delta(\bm{k})$ in terms of the effective pairing interaction $\tilde{U}(\bm{q})$, where $\bm{q}=\bm{k}-\bm{k'}$.
In Figs. \ref{f2}(b)-\ref{f2}(d), we show  $\tilde{U}(\bm{q})$, $\chi_{\rm s}(\bm{q})$, and $\chi_{\rm c}(\bm{q})$. 
We find that  $\tilde{U}(\bm{q})$ is negative in the entire $\bm{q}$ region, meaning that the effective interaction between color-1 and -2 atoms is attractive. We also find a large attractive peak at $\bm{q}\sim (\pi, \pi)$. Note that $\tilde{U}(\bm{q})$ has peaks at $\bm{q} \sim (\pm\pi, \pm\pi)$ and $(\pm\pi, \mp\pi)$. 
To gain these strong attractive interactions at $\bm{q} \sim (\pm\pi,\pm\pi)$ and $(\pm\pi,\mp\pi)$, $\Delta(\bm{k})$ has the large amplitude at $\bm{k} \sim (\pm\pi,0)$ and $(0,\pm\pi)$.

The origin of the strong attractive interaction can be explained in terms of  $\chi_{\rm s}(\bm{q})$ and $\chi_{\rm c}(\bm{q})$. As shown in Eq. (\ref{effU}),  $\tilde{U}(\bm{q})$ is described by two susceptibilities, $\chi_{\rm s}(\bm{q})$ and $\chi_{\rm c}(\bm{q})$. 
Figures \ref{f2}(c) and \ref{f2}(d) show that both $\chi_{\rm s}(\bm{q})$ and $\chi_{\rm c}(\bm{q})$ have the peaks at $\bm{q}  \sim (\pi, \pi)$, and $\chi_{\rm c}(\bm{q})$ overcomes $\chi_{\rm s}(\bm{q})$ in the entire $\bm{q}$ region, leading to the large attractive peak in $\tilde{U}(\bm{q})$ at $\bm{q} \sim (\pi, \pi)$. 
As shown in Fig. \ref{f3}(a), the peaks in $\chi_{\rm c}(\bm{q})$ as well as $\tilde{U}(\bm{q})$ develop notably with decreasing temperature, while the peak in $\chi_{\rm s}(\bm{q})$ is almost independent of temperature. 
These results indicate that the large attractive peak in $\tilde{U}(\bm{q})$ is caused by $\chi_{\rm c}(\bm{q})$.

Since the momentum $\bm{q}=(\pi, \pi)$ describes the perfect nesting condition in square optical lattices at half filling, we consider that the peak in $\chi_{\rm c}(\bm{q})$ at $\bm{q} \sim (\pi, \pi)$ is caused by quantum fluctuations of the CDW that appears at half filling for $U/U'<1$ \cite{Miyatake2010,Inaba2013}. 
To confirm this consideration, we first investigate the transition temperature $T_{\rm c}$ by changing filling $N$. As shown in Fig. \ref{f1}(b), $T_{\rm c}$ increases with increasing $N$ towards half filling $N=3/2$. This is because as $N$ is increased CDW fluctuations are enhanced. 
In Fig. \ref{f3}(b), we show the filling  dependencies of $\tilde{U}(\bm{q})$, $\chi_{\rm s}(\bm{q})$, and $\chi_{\rm c}(\bm{q})$ for $U/U'=0.01$ and $U'/t=1.2$ at $T \sim1.2T_{\rm c}$. 
The peaks develop definitely in $\chi_{\rm c}(\bm{q})$ and $\tilde{U}(\bm{q})$ with $N$ approaching half filling.  
Note that for $N<1.318$ the extended $s$-wave superfluid state disappears, while for  $N>1.362$ our diagrammatic method fails to calculate the susceptibilities  $\chi_{\rm s}(\bm{q})$ and $\chi_{\rm c}(\bm{q})$.

Since the order parameter has a full gap, the extended $s$-wave superfluid gap consists of a superposition of the $\bm{k}$-independent uniform component and the $\bm{k}$-dependent component.  
We consider that the former component is caused by local density fluctuations of unpaired atoms \cite{Inaba2012} and the latter one is caused by nonlocal CDW fluctuations. 
Thus, the extended $s$-wave superfluid state is characteristic of the three-component repulsive lattice fermionic atoms and appears nontrivially in spite of a single-band repulsive lattice fermions at close to half filling.

\subsection{Two weaker repulsions than the other}

\begin{figure}
\begin{center}
\includegraphics[width=\linewidth]{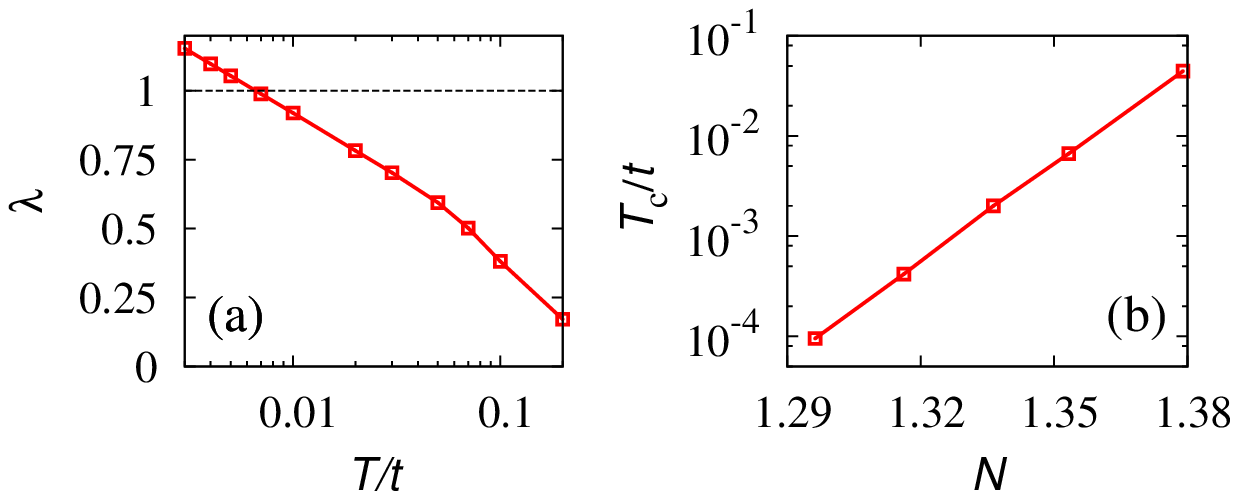}
\end{center}
\caption{(Color online) (a)  (a) Eigenvalue $\lambda$ as a function of temperature $T$ for $U/U'=1.3$, $U'/t=1.2$, and $N \sim1.35$. (b) Transition temperature $T_{\rm c}$ as a function of filling $N$ for $U/U'=1.3$ and $U'/t=1.2$. 
}
\label{f4}
\end{figure}
\begin{figure}
\begin{center}
\includegraphics[width=\linewidth]{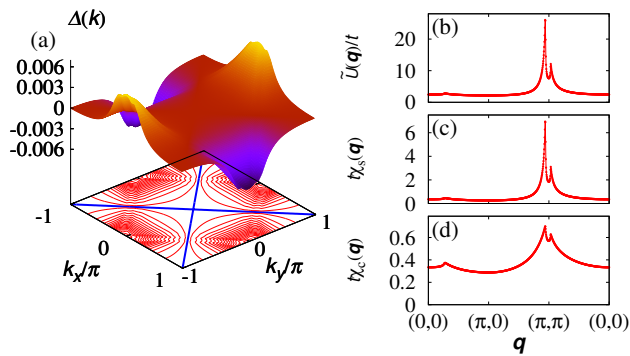}
\end{center}
\caption{(Color online)  (a) Superfluid order parameter $\Delta(\bm{k})$, (b) effective pairing interaction $\tilde{U}(\bm{q})$, and susceptibilities (c) $\chi_s(\bm{q})$ and (d) $\chi_c(\bm{q})$ for $U/U'=1.3$, $U'/t=1.2$, $N \sim1.35$, and $T/t=0.007$,  which is slightly higher than $T_{\rm c}/t \sim 0.0066$. Momenta satisfy $\bm{q}=\bm{k}-\bm{k'}$. Thick (blue) lines in $k_x$-$k_y$ plane represent nodes.  
}
\label{f5}
\end{figure}
\begin{figure*}[htb]
\begin{center}
\includegraphics[width=130mm]{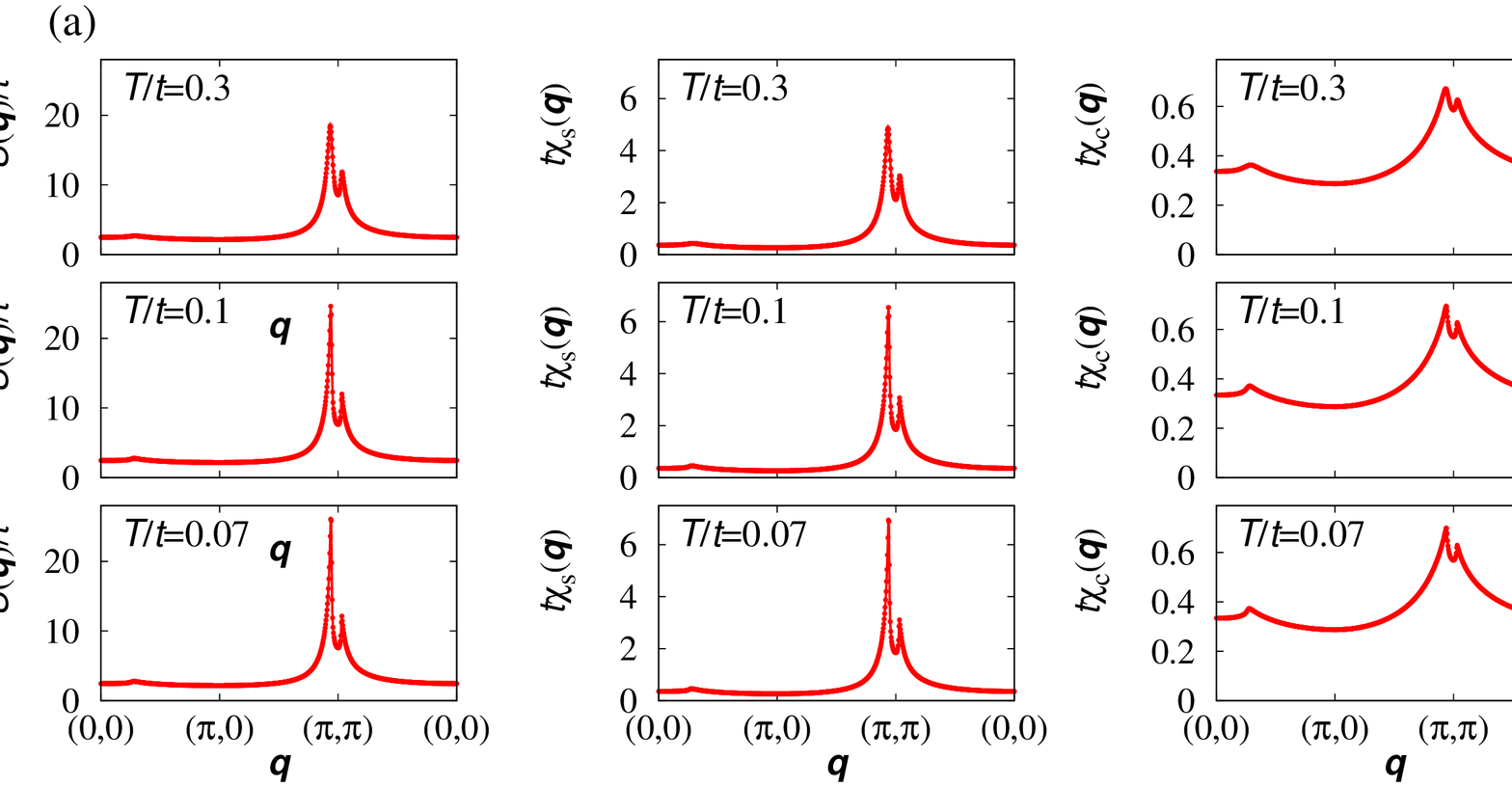}
\includegraphics[width=130mm]{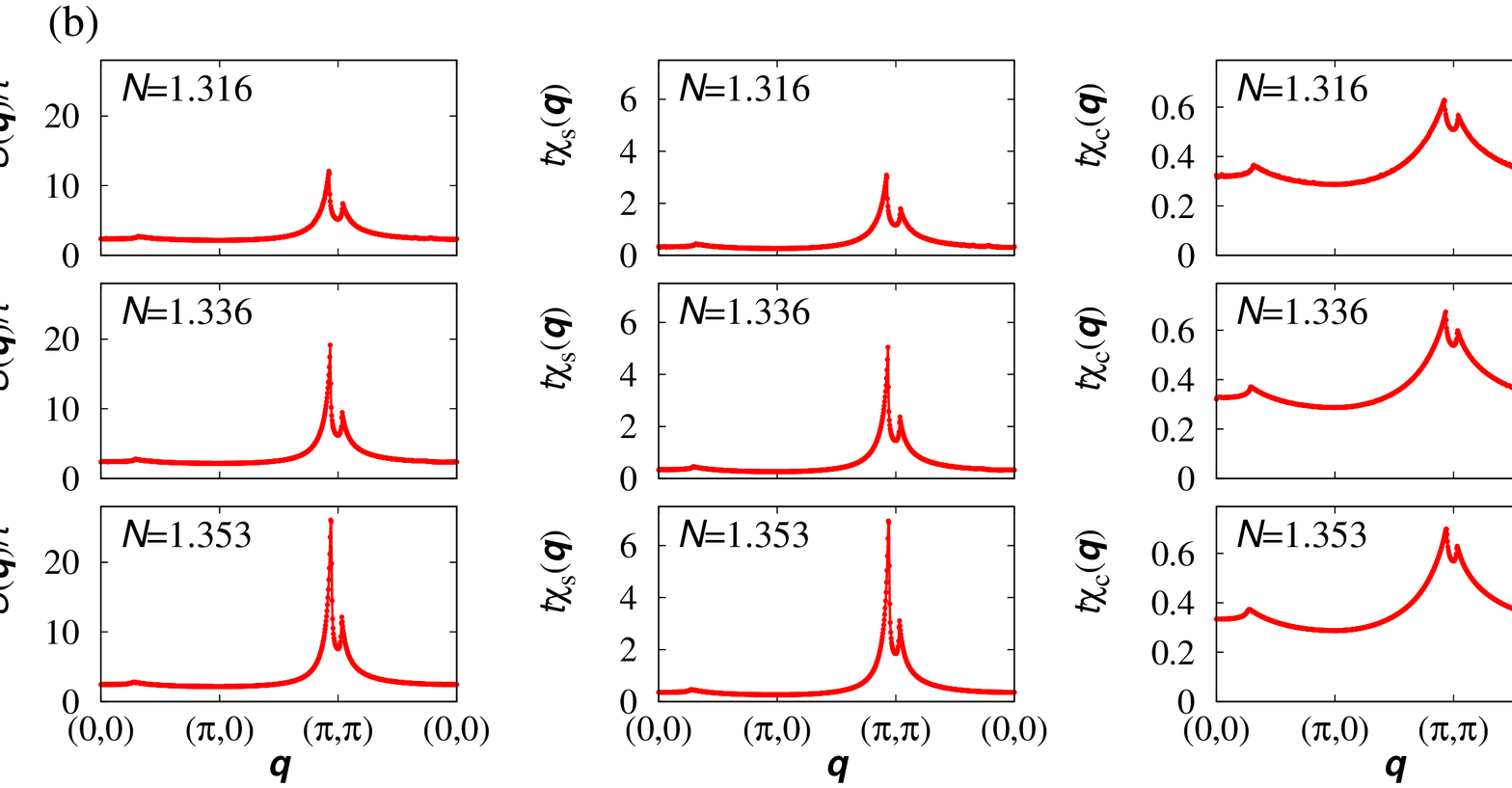}
\end{center}
\caption{(Color online)  (a) Temperature dependencies of effective interaction $\tilde{U}(\bm{q})$, susceptibilities $\chi_s(\bm{q})$ and  $\chi_c(\bm{q})$ for $U/U'=1.3$, $U'/t=1.2$, and $N \sim1.35$. 
(b) Filling dependencies of effective interaction $\tilde{U}(\bm{q})$, susceptibilities $\chi_s(\bm{q})$ and  $\chi_c(\bm{q})$ for $U/U'=1.3$, $U'/t=1.2$, and $T\sim 1.2T_{\rm c}$.  }
\label{f6}
\end{figure*}
We next investigate the superfluid state, when the difference in the repulsive interactions is reduced ($U/U'$ is increased). 
Figure \ref{f4}(a) shows the eigenvalue $\lambda$ as a function of $T$ for $U'/t=1.2$, $U/U'=1.3$, and $N \sim1.35$. We evaluate $T_{\rm c}/t \sim 0.0066$. The order parameter $\Delta(\bm{k})$ at $T/t=0.007$ is shown in Fig. \ref{f5}(a). 
We find two nodal lines, $k_y=\pm k_x$, indicating that the pairing symmetry is a $d_{x^2-y^2}$ wave.
As shown in Figs. \ref{f5}(b)-\ref{f5}(d) $\chi_{\rm s}(\bm{q})$ overcomes $\chi_{\rm c}(\bm{q})$, which yields a repulsive $\tilde{U}(\bm{q})$ in the entire $\bm{q}$ region with a notable peak at $\bm{q}\sim (\pi,\pi)$. 
This situation is in contrast to the extended $s$-wave superfluid state. 
In Figs. \ref{f6}(a) and \ref{f6}(b), we show the $T$ and $N$ dependencies of $\tilde{U}(\bm{q})$, $\chi_{\rm s}(\bm{q})$, and $\chi_{\rm c}(\bm{q})$, respectively. 
The peaks in $\chi_{\rm s}(\bm{q})$ and $\tilde{U}(\bm{q})$ develop conspicuously at $\bm{q} \sim (\pi, \pi)$ with decreasing $T$ [Fig. \ref{f6}(a)] or increasing $N$ towards half filling [Fig. \ref{f6}(b)]. On the other hand, the peak in $\chi_{\rm c}(\bm{q})$ remains almost unchanged. We also find that $T_{\rm c}$ increases with increasing $N$, as shown in Fig. \ref{f4}(b). 
Since $\bm{q}=(\pi,\pi)$ describes the perfect nesting condition for the CSAF that appears at half filling for $U/U'>1$ \cite{Miyatake2010,Inaba2013}, the results indicate that the large repulsive peak in $\tilde{U}(\bm{q})$ is caused by quantum fluctuations of the CSAF.  
We thus conclude that the CSAF fluctuations induce the $d_{x^2-y^2}$-wave superfluid state. 


\subsection{Phase diagram and discussions}

\begin{figure}
\begin{center}
\includegraphics[width=0.9\linewidth]{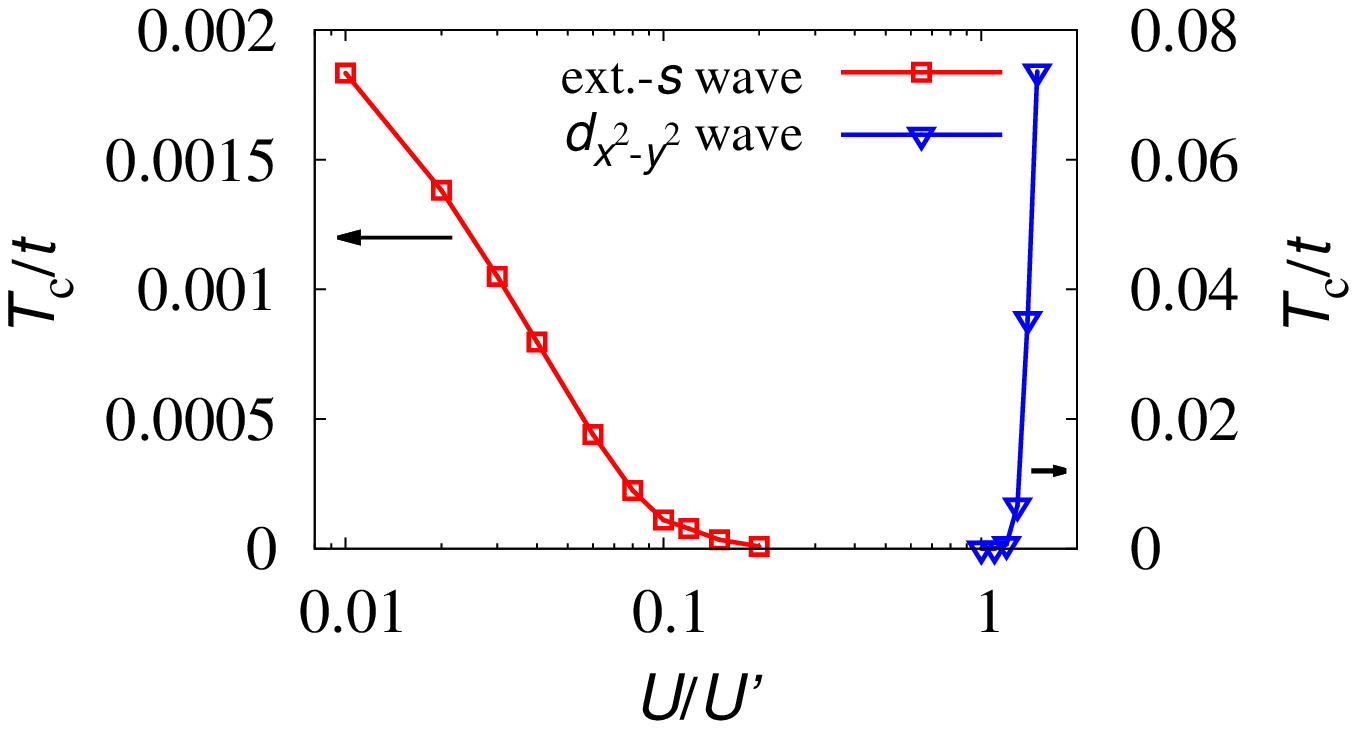}
\end{center}
\caption{(Color online) Pairing symmetry and transition temperature $T_{\rm c}$ as a function of $U/U'$ for $U'/t=1.2$ and $N \sim1.35$. }
\label{f7}
\end{figure}
We perform the same calculations by changing $U/U'$ for $U'/t=1.2$ and $N \sim1.35$, and evaluate $T_{\rm c}$. We summarize the results in the phase diagram shown in Fig. \ref{f7}. 
The extended $s$-wave superfluid state appears in $U/U' \lesssim 0.2$ and $T_{\rm c}$ increases with decreasing $U/U'$. The extended $s$-wave pairing is caused by the cooperation of the local density fluctuations of unpaired atoms and the nonlocal CDW fluctuations
Recently, $s$-wave superconductivity driven by local spin fluctuations was found in the paramagnetic sector of the Kondo lattice model \cite{Bodensiek}. In this study, a question arose concerning the pairing symmetry driven by the competition or cooperation between local and nonlocal fluctuations. Our results provide an answer to this question. 
On the other hand, the $d_{x^2-y^2}$-wave superfluid state appears in $U/U'>1$ and $T_{\rm c}$ increases with increasing $U/U'$.

The present results capture the essentials of the superfluid state in repulsively interacting three-component fermionic atoms in optical lattices. When the higher-order corrections of correlation effects are included and/or the \'{E}liashberg equation is derived within a strong-coupling theory, we consider that the quantitative features of the phase diagram are refined but the qualitative features remain.

\section{\label{sec:level4}Experimental Observability and Summary}
We discuss the relationship between our theoretical results and the experiments.
As discussed before, a three-component $^6{\rm Li}$ fermionic gas has been realized \cite{Ottenstein2008,Huckans2009}. 
As shown in Fig. 2(c) in Ref. \cite{Ottenstein2008}, the magnetic Feshbach resonance of $^6{\rm Li}$ allows us to control the color dependent interactions. The repulsive region has been observed under a magnetic field from 550 to 630G with a very small three-body loss \cite{Ottenstein2008}. 
At close to 600G two repulsions are much stronger than the other, while at close to 630G two repulsions are weaker than the other. 
We thus expect the three-component $^6{\rm Li}$ atoms in optical lattices to be a possible candidate for observing the change in the pairing symmetry.
Another candidate is a selective mixture of $^{171}{\rm Yb}$ and $^{173}{\rm Yb}$ fermionic atoms in optical lattices. In this system, the color-dependent interactions are naturally induced and can be controlled by the optical Feshbach resonance \cite{Enomoto2008}.

State-of-the-art experimental techniques such as rf-spectroscopy \cite{Stewart2008} can allow us to detect momentum-resolved single-particle excitation spectra. This quantity provides useful information for detecting a superfluid transition and its pairing symmetry. 
We discuss the features of momentum-resolved single-particle excitation spectra at the special momenta $\bm{k}=(0,0)$, $(\pi/2,\pi/2)$, and $(\pi,0)$. 
In the extended $s$-wave pairing, the full spectral gap opens. On the other hand, in the $d_{x^2-y^2}$-wave pairing, the spectral gap closes at $\bm{k}\sim(0,0)$ and $(\pi/2,\pi/2)$, while the spectral gap opens at $\bm{k}\sim(\pi,0)$.
Thus the superfluid gap of each pairing symmetry exhibits characteristic behavior at these three momenta, which makes it possible to probe the change in pairing symmetry experimentally.

In summary, we have shown that three-component repulsive fermionic atoms in optical lattices can act as a quantum simulator for controlling the pairing symmetry of superfluid states.
We hope to confirm this fascinating phenomenon experimentally.

\begin{acknowledgments}
We would like to thank K. Inaba for fruitful comments and valuable discussions. 
We also thank A. Koga and H. Tsuchiura for useful discussions.
We acknowledge the numerical resource provided by the Research Center for Nano-micro Structure Science and Engineering at University of Hyogo. 
This work was supported by JSPS KAKENHI Grants No. 25287104 and No. 15K05232. 
\end{acknowledgments}

\end{document}